\documentclass[fleqn,twoside]{article}
\usepackage{espcrc2}
\usepackage{epsfig}

\usepackage{graphicx}
\usepackage[figuresright]{rotating}

\newcommand{\AmS}{{\protect\the\textfont2
  A\kern-.1667em\lower.5ex\hbox{M}\kern-.125emS}}

\newcommand{\be}{\begin{equation}}
\newcommand{\ee}{\end{equation}}
\newcommand{\ba}{\begin{eqnarray}}
\newcommand{\ea}{\end{eqnarray}}

\newcommand{\ii}{{\rm i}}

\newcommand{\op}{{\cal O}}

\newcommand{\RR}{{\rm I\kern -.2em  R}}
\def\lsi{\raise0.3ex\hbox{$<$\kern-0.75em\raise-1.1ex\hbox{$\sim$}}}
\def\gsi{\raise0.3ex\hbox{$>$\kern-0.75em\raise-1.1ex\hbox{$\sim$}}}
\newcommand{\lsim}{\mathop{\lsi}}

\title{QCD phase diagram for small densities from simulations at imaginary $\mu$}

\author{P.~de Forcrand\address{ETH, CH-8093 Z\"urich, Switzerland and
CERN Theory Division, CH-1211 Geneva 23, Switzerland}
        and
        O.~Philipsen\address{Center for Theoretical Physics, MIT, 
Cambridge, MA 02139-4307, USA}
\thanks{Talk given by O.~P.}}
       
\begin{document}

\begin{abstract}
We present results on the QCD phase diagram for small densities without
reweighting.
Our simulations are performed with an imaginary chemical potential $\mu_I$
for which the fermion determinant is positive.
On an $8^3\times 4$ lattice with 2 flavors of staggered quarks,
we map out the pseudo-critical temperature $T_c(\mu_I)$.
For $\mu_I/T \leq \pi/3$, this is an analytic function
whose Taylor expansion converges rapidly, with truncation errors smaller than
statistical ones. The result is analytically continued to give the location of the pseudo-critical
line for real $\mu_B\lsim 500$ MeV. 
\vspace{1pc}
\end{abstract}

\maketitle

\section{INTRODUCTION}

In view of heavy ion collision experiments a pressing task
for lattice QCD is to map out the deconfinement transition 
in the $(\mu,T)$-plane.
The complexity of the fermion determinant at finite baryon density renders
standard Monte Carlo techniques impossible \cite{rev}.
One way of circumventing this `sign problem' are multi-dimensional reweighting techniques,
and a phase diagram has been presented last year \cite{fk2}. 
Reweighting generally breaks down for large volumes and/or densities, 
and one difficulty 
of the approach is to know when that happens. 
In \cite{hk} both reweighting factor and observables were expanded
in a Taylor series in $\mu/T$, and the first coefficients were measured \cite{hk}.
This method allows to simulate larger volumes, but a priori
nothing is known about the error introduced by truncating the series.

Here we present an alternative method avoiding reweighting altogether.
We simulate at imaginary $\mu$ for which there is no sign problem and
arbitrarily large volumes are feasible.
Observables are fitted by truncated Taylor series in $\mu/T$, keeping full control
over the associated systematic error. The series may then be
continued to real $\mu$, provided the observable is analytic \cite{lom}.
This strategy has been successfully tested for screening masses \cite{hlp2},
here we extend it to the critical line itself. A detailed account of this work 
is given in \cite{us}.

\begin{figure}[tbh]
\centerline{\epsfxsize=7cm\hspace*{0cm}\epsfbox{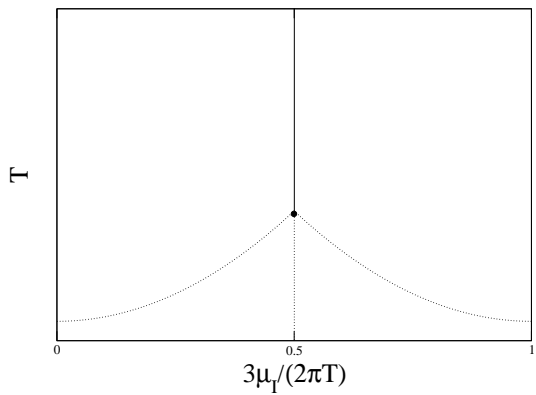}}
\vspace*{-0.8cm}
\caption[a]{\label{schem}
Phase diagram in the $(\mu_I,T)$-plane.
The vertical line marks a $Z(3)$ transition crossing
the deconfinement transition. For $T>T_c$ the former is of first order.
}
\end{figure}
\section{QCD AT IMAGINARY $\mu$}

\begin{figure}[th]
\centerline{\epsfxsize=6cm\hspace*{0cm}\epsfbox{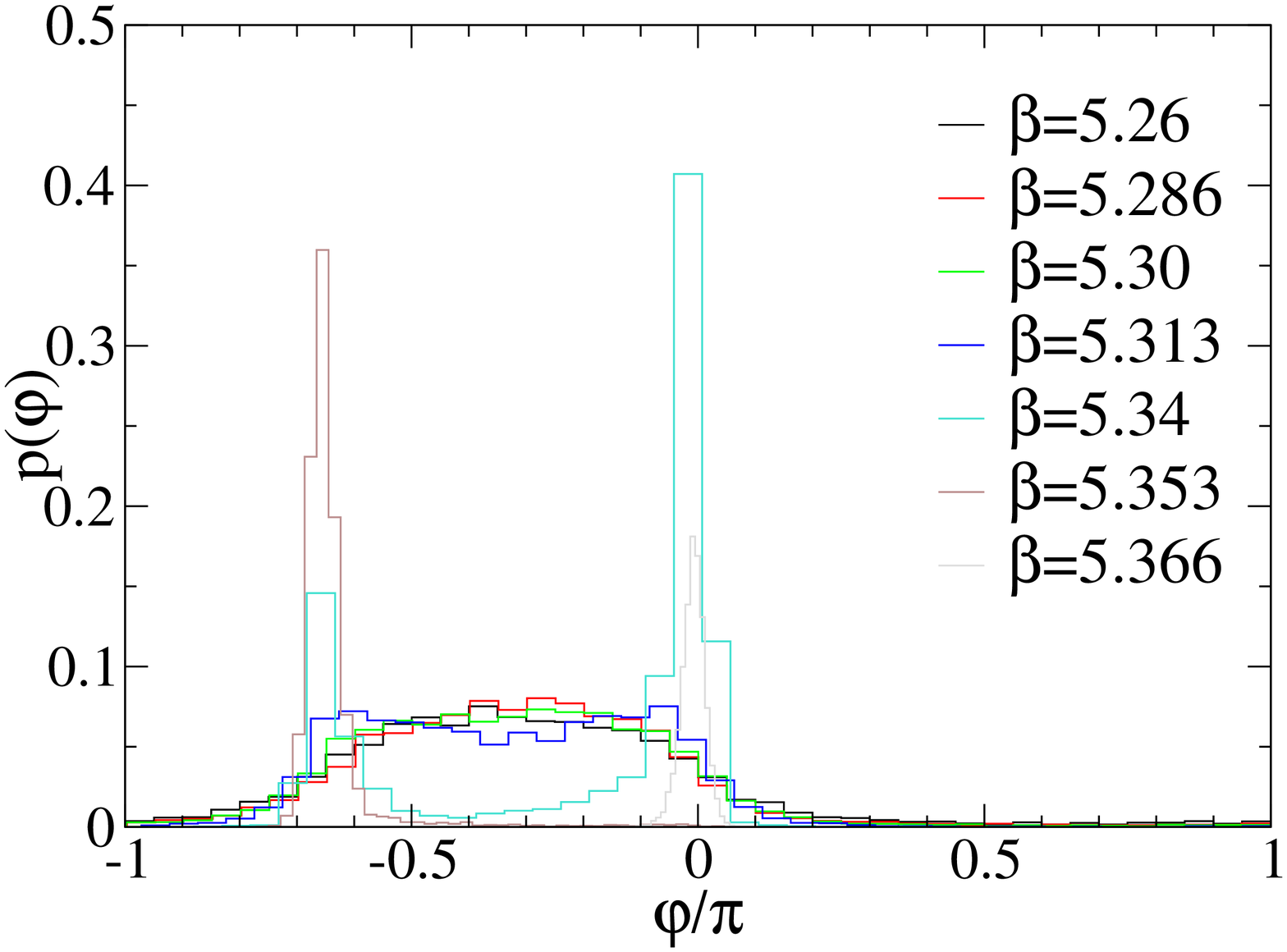}}
\vspace*{-0.8cm}
\caption[a]{
Probability distribution of the phase of the Polyakov loop for the critical
value $a\mu_I^c=\pi/12$.
}
\label{phaseP_hist}
\vspace*{-0.5cm}
\end{figure}
Finite density QCD is symmetric under
reflection of $\mu$, $Z(\mu)=Z(-\mu)$, for both real and imaginary $\mu$. 
For imaginary $\mu={\rm i}\mu_I$
it is moreover periodic, $Z(\mu_I)=Z(\mu_I+2\pi/3)$ \cite{weiss}. 
Hence, certain shifts in 
$\mu_I$ are equivalent to Z(3) transformations to different vacua.
Consequently, Z(3) transitions occur at the critical values 
$(\mu/T)^c=(k+1/2)2\pi/3$. The Z(3) sectors can be identified by monitoring the phase
of the Polyakov loop. Ref.~\cite{weiss} predicts these transitions to be of
first order for $T>T_c$ and crossover for $T<T_c$, with the critical
deconfinement temperature $T_c$. Schematically, the 
phase diagram is shown in Fig.~\ref{schem}, which is periodically repeated for
larger $\mu_I$.

Our simulations support this picture. Fig.~\ref{phaseP_hist} shows
the phase of the Polyakov loop at $(\mu/T)^c$ for 
various lattice couplings/temperatures. A smooth distribution is found for low
temperatures, whereas at high temperatures the two-state distribution signals 
a first order phase transition. Similar findings are reported for four flavors
in \cite{it}. The existence of a Z(3) transition with its endpoint joining on 
the deconfinement transition implies that the critical line of the latter becomes
non-analytic at that point, and hence only $\mu_I\leq \pi/3$ can be continued to
real $\mu$. 

Note that the deconfinement line as well as all observables even under reflections
$\mu\rightarrow -\mu$ are symmetric around the Z(3) transition. This is a feature
that is not reproduced by reweighting from a Monte Carlo ensemble
generated at $\mu=0$, since
such an ensemble does not sufficiently probe the other Z(3) sectors
\cite{us,fk2}.
\begin{figure}[th]
\centerline{\epsfxsize=6cm\hspace*{0cm}\epsfbox{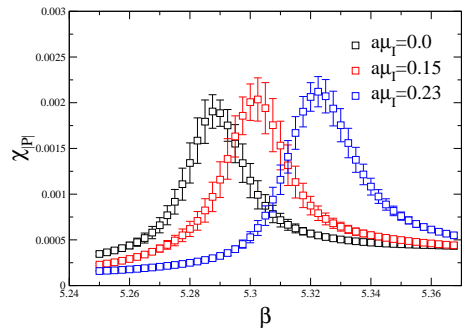}}
\vspace*{-0.8cm}
\caption[a]{\label{susc}
Susceptibilities of the modulus of the Polyakov loop for various $\mu_I$.
}
\end{figure}

\section{CONTINUATION OF THE DECONFINEMENT LINE}

One method to locate a phase transition is to look for the peak
$\chi_{max}=\chi(\mu_c,\beta_c)$
of the susceptibilities $\chi= VN_t \left\langle(\op - \langle\op\rangle)^2\right\rangle$,
where we have used the plaquette, the chiral condensate and the Polyakov loop as
operators $\op(x)$.
Since simulations are always performed on finite volumes, susceptibilities are analytic
functions over the whole parameter space of the theory. 
Non-analyticities associated with phase transitions develop only in the thermodynamic
limit.

For a given finite volume, the pseudo-critical line
is then found by locating $\chi_{max}$ for every $\mu$,
\be \label{crit}
\left.\frac{\partial\chi}{\partial\beta}\right|_{\beta_c}=0\quad
\left.\frac{\partial^2\chi}{\partial\beta^2}\right|_{\beta_c}<0.
\ee
This is an implicit definition of the critical coupling $\beta_c(\mu)$.
For analytic $\chi(\mu,\beta)$ the implicit function theorem implies that also
$\beta_c(\mu)$ is analytic for all $\mu$. 
Moreover, the symmetry considerations of the last section 
ensure $\beta_c(\mu)=\beta_c(-\mu)$, so that we can expand in $\mu/T=aN_t\mu$ to obtain
\be \label{beta}
\beta_c(\mu)=\sum_{n}c_n (a\mu)^{2n}.
\ee
As long as, for imaginary $\mu={\rm i}\mu_I$, the non-perturbative data are well described by 
a few coefficients in this series, analytic continuation to real $\mu$ simply proceeds
by the operation $\mu_I\rightarrow \ii \mu_I$. This provides us with the location
of the pseudo-critical line on a finite volume.

If there is a first order deconfinement transition ending in a critical point,
then, as the volume is increased, $\beta_c(\mu)$  approaches the line of phase transitions,
while remaining pseudo-critical in a crossover regime.
In principle, the nature of the line can be determined from the way
the thermodynamic limit is approached, 
$(\beta_c(V)-\beta_c(\infty))\sim V^{-\sigma}$, with $\sigma=1$ or $0<\sigma<1$ 
for a first
or second order transition, respectively, and $\sigma=0$ for a crossover.
Our first simulations are on a single volume and this question is postponed to 
future work.

Fig.\ref{susc} shows a sequence of susceptibilities for the modulus of the Polyakov 
loop, with similar pictures for the plaquette and the chiral condensate.
All three observables give statistically compatible values $\beta_c(\mu_I)$.
Note that the peak is moving to larger $\beta_c$ for growing $\mu_I$, as one would expect
for imaginary $\mu$, cf.~Eq.~\ref{beta} and Fig.~\ref{schem}.

\begin{figure}[t]
\vspace*{-0.2cm}
\centerline{\epsfxsize=6cm\hspace*{0cm}\epsfbox{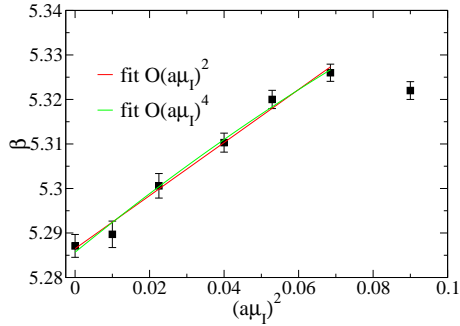}}
\vspace*{-0.8cm}
\caption[a]{\label{pdiag}
Location of the pseudo-critical deconfinement line as determined from plaquette
susceptibilities. The lines show fits of the form Eq.~\ref{beta} to quadratic and
quartic order.
}
\end{figure}
Fig.~\ref{pdiag} shows $\beta_c$ for different values
of $\mu_I$. The data perfectly reproduce the expected symmetry of the partition 
function about the first Z(3) transition point.
The lines in the figure show fits of a quadratic and a quartic polynomial to the data,
with $\chi^2/{\rm dof}\approx$0.6 and 0.7, respectively. The coefficient of the quartic
term is found to be zero within errors, and hence the two fits are statistically 
compatible: up to $\mu_I/T=\pi/3$, and for the quark flavors and
masses under consideration, the critical line is well described by the quadratic
term only.

\begin{figure}[t]
\vspace*{-0.2cm}
\centerline{\epsfxsize=6cm\hspace*{0cm}\epsfbox{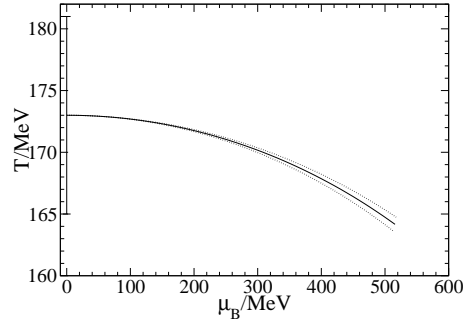}}
\vspace*{-0.8cm}
\caption[a]{
Location of the deconfinement transition.
The error bar gives the uncertainty
in $T_c(0)$ used to set the scale, the dotted
lines reflect the error on $c_1$, Eq.~\ref{beta}.
}
\label{rpdiag} 
\end{figure}

After analytic continuation 
the result is transformed into physical units by means of the perturbative two-loop 
expression for the scale parameter $a\Lambda_L$ and using $T_c(\mu=0)$ from the literature \cite{kar} 
to set the physical scale. We thus arrive at the phase diagram of Fig.~\ref{rpdiag}, with 
the pseudo-critical line given by ($\mu_B=3\mu$)
\be
\frac{T_c(\mu_B)}{T_c(\mu_B=0)}= 1 - 0.00563(38) \left(\frac{\mu_B}{T}\right)^2.
\ee
This result is consistent with those obtained by the reweighting approaches \cite{fk2,hk},
for which it provides a crucial, controlled check.
An important task ahead is to perform
a finite size scaling analysis of the continued critical line,
in order to determine the order of the transition as well as to locate
a possible critical point. 

\end{document}